\documentclass[english,aps,manuscript]{revtex4}
\usepackage[T1]{fontenc}
\usepackage[latin1]{inputenc}
\usepackage{amssymb}

\makeatletter

\newcommand{\bee}{\begin{equation}}
\newcommand{\ee}{\end{equation}}
\newcommand{\beea}{\begin{eqnarray}}
\newcommand{\eea}{\end{eqnarray}}

\usepackage{babel}
\makeatother
\begin{document}

\preprint{COLO-HEP-526}

\maketitle
\begin{center}\textbf{\Large On Gauge Mediated SUSY Breaking and Moduli
Stabilization }\par\end{center}{\Large \par}

\begin{center}\vspace{0.3cm}\par\end{center}

\begin{center}{\large S. P. de Alwis$^{\dagger}$ }\par\end{center}{\large \par}

\begin{center}Physics Department, University of Colorado, \\
 Boulder, CO 80309 USA\par\end{center}

\begin{center}\vspace{0.3cm}\par\end{center}

\begin{center}\textbf{Abstract}\par\end{center}

\begin{center}\vspace{0.3cm}\par\end{center}

A generic lesson of string theory is that the coupling constants of
an effective low energy theory are determined by the vacuum values
of a set of fields - the so-called moduli - some of which are stabilized
at relatively low masses by non-perturbative effects. We argue that
the physics of these moduli cannot be separated from the issues of
dynamical and gauge mediated supersymmetry breaking. To illustrate
this point we present a modified version of the type IIB KKLT model
where the criteria for gauge mediated SUSY breaking may be realized.

\vspace{0.3cm}


\vfill

$^{\dagger}$ {\small e-mail: dealwiss@colorado.edu}{\small \par}

\eject

\section{Introduction}

\subsection{String theory and SUSY breaking}

There are two possible ways in which string theory may have relevance
for physics at the Large Hadron Collider (LHC) scales of 1-2 TeV. 

\begin{itemize}
\item The effective string scale is very low, either because we live on
a brane which is situated far down some strongly warped throat or
because the compactification volume is very large. In this case one
would expect to see spectacular signatures of an underlying stringy
world at these scales.
\item The second possibility is that the string solution describing our
world has a low scale of supersymmetry (SUSY) breaking with mass splittings
on the order of a few hundred GeV, in which case we should see superpartners
of the standard model particles. 
\end{itemize}
The first possibility - if it is realized in our universe - will almost
certainly lead to a confirmation of the underlying stringy nature
of fundamental physics. However it appears to us to be the less likely
one. After all the one piece of evidence that we currently have for
the existence of physics beyond the standard model, is the apparent
unification of the couplings at a scale of around $10^{16}GeV$, if
we include the threshold effects of superpartners of the standard
model at a scale of around 1TeV. In addition low energy supersymmetry
breaking seems to be related to the spontaneous breakdown of the standard
model symmetry in an elegant way. Of course the observation of superpartners
(unlike say the observation of black holes) at the LHC scale will
not directly support string theory. Nevertheless it will tell us that
we are headed in the right direction, and furthermore will give us
restrictions (apart from requiring 4 dimensions, a small cosmological
constant, three generations, etc.) on the landscape of solutions that
may describe our world. 

Generically string theory prefers to have an AdS SUSY minimum or one
with SUSY broken at a high (i.e. string) scale. Since both of these
are phenomenologically irrelevant we have to restrict ourselves to
those (non-generic) models which have a low SUSY breaking scale %
\footnote{Of course the same problem exists (perhaps in a even more acute form)
for the large volume (or large warping) solutions. %
}. So we assume at the outset that the LHC will find superpartners
of the standard model so that we will be restricted to those models
which give a low SUSY breaking scale.

We will focus on solutions with small or moderate warping and moderately
large extra dimensions, i.e. the Kaluza-Klein masses $m_{KK}<M_{s}$
(but not $<<M_{s}$) and $M_{s}<M_{P}$ (but not $<<M_{P}$). In string
theory generic compactifications should have moduli stabilized at
or near the string scale. However such situations are hard to deal
with in practice. For instance the recent progress made in understanding
how the moduli are stabilized (see \cite{Grana:2005jc} and \cite{Douglas:2006es}
for recent reviews) depends on a starting point - namely ten-dimensional
supergravity (SUGRA) - that assumes that there is a clear separation
between the string scale and the scale of the compactification. On
the other hand if we want to preserve unification then the theory
should be four dimensional upto a scale of $10^{16}GeV$ (which is
also around the most likely scale of inflation). This translates into
having an internal manifold whose characteristic size is around two
orders of magnitude larger than the Planck size and an order of magnitude
larger than the string size.

We will proceed as usual by arguing that the theory can be discussed
within the 4D SUGRA framework. As in the work of KKLT \cite{Kachru:2003aw}
if one can find a stabilization which leads to a compactification
manifold whose size is in the above range the assumption is at least
self-consistent. Then we need to address how supersymmetry is broken
in the theory and whether it is possible to generate a small scale
of SUSY breaking. Finally one has to have a mechanism to transmit
the SUSY breaking to the visible (MSSM) sector.

SUSY breaking is communicated to the visible sector in three possible
ways. Setting the Planck scale $M_{P}=1$ the SUSY breaking order
parameter is an F-term and/or a D-term with $0\ne F|_{0}<<1$ (and/or
$0\ne D|_{0}<<1$). Typically one expects $F_{0}\sim m_{3/2}\le O(10^{-15})$. 

\begin{enumerate}
\item Moduli/gravity mediation. This usually involves SUSY breaking in the
moduli sector (for example $F_{T}\ne0$). The moduli sector couples
then to the visible sector though gravitational strength interactions.
However it may still be the dominant mechanism even if SUSY is broken
in some (hidden) matter sector.
\item Anomaly mediated. $F_{\varphi}\ne0,$ where $\varphi$ is the conformal
(or Weyl) compensator of SUGRA.
\item Gauge mediated. In this case one needs a hidden matter sector $\phi$
with $F_{\phi}\ne0$. The supersymmetry breaking is then transmitted
through gauge interactions of the so-called messengers which are directly
coupled to $\phi$.
\end{enumerate}
1. is generically plagued with the SUSY flavor changing neutral current
(FCNC) problem as well as the SUSY CP problem, though there are special
situations where (at least classically) the problem may be avoided.
Even in those, quantum effects will usually generate FCNC. 2. by itself
would not have the FCNC problem but leads to tachyonic sleptons. In
any case it occurs along with 1. (see for example \cite{Choi:2005ge}
and \cite{Dine:2005iw}) and it is hard to suppress 1. while keeping
2 and avoiding tachyonic sleptons. On the other hand 3. will solve
the FCNC problem since gauge interactions are flavor neutral. The
question is whether this property will survive the embedding of this
mechanism in a string theory with all moduli stabilized. 

We need to understand the minimum requirements on the string theory
model which will enable the gauge mediated mechanism to be the dominant
one. It is important to analyze the problem within a supergravity
(SUGRA) or string theory context. Dynamical breaking with metastable
vacua necessarily involve (at least) two distinct minima with different
cosmological constants and this situation cannot really be discussed
within a global SUSY context as has been done upto now in almost all
discussions - for a recent discussion see \cite{Intriligator:2007cp}.
Whithin a SUGRA context general conditions on F-terms in the supersymmetry
breaking sector have been discussed in \cite{Brustein:2001ci}. In
the particular case of KKLT type models \cite{Kachru:2003aw}, the
difficulties involved in uplifting the minimum of the potential to
get a small positive cosmological constant with supersymmetry breaking
at a parametrically larger scale has been discussed in \cite{Brustein:2004xn},
where it was also pointed out that one can get F-term uplifiting if
the Kaehler potential is changed from the simple KKLT form. Also as
has been discussed in \cite{deAlwis:2006nm} tuning the CC in the
flux compactified context affects all other parameters (Yukawa couplings
soft masses etc.) and it is only within the context of the SUGRA formalism
that a careful discussion of what predictions can be extracted from
a given class of models be made. 

The main focus of this paper will be how the mechanism of moduli stabilization
in string theory can be compatible with gauge mediated SUSY breaking.

\subsection{Gauge mediated SUSY breaking }

Gauge mediated SUSY breaking (for reviews see \cite{Luty:2005sn}\cite{Giudice:1998bp})
has a symmetry breaking sector (here simplified to the field $\phi$)
and a vector like messenger sector (which is charged under the standard
model group) here denoted by $f,\tilde{f}$. The latter may be taken
as say a $5+\bar{5}$ of the $SU(5)$ which contains the standard
model group so that gauge coupling unification is not affected. The
relevant superpotential term is $W=\phi f\tilde{f}$. The superfield
$\phi$ is then supposed to get both a vacuum expectation value and
a non-zero F-term. In other words $\phi_{0}=M+\theta^{2}F$ where
$\theta$ is the fermionic superspace coordinate. This then causes
a mass splitting between the fermionic and the bosonic components
of the messenger fields $\Delta m_{f}^{2}=F$. The messengers couple
to the standard model fields only through gauge couplings (essentially
through threshold effects in the running of the gauge coupling) and
so communicate the SUSY breaking in a flavor neutral way. The gaugino
masses are then given by \begin{equation}
\frac{m_{\lambda}(\mu)}{g^{2}(\mu)}=\frac{N}{(4\pi)^{2}}\frac{F}{M}\label{eq:gagino}\end{equation}
 where $\mu$ is the RG scale and $N$ is the number of messengers
(assumed to be in the fundamental representation) at the mass scale
$M$. This is indeed an elegant and relatively simple solution to
the flavor problem provided the following issues are addressed.

\begin{itemize}
\item What is the mechanism of SUSY breaking? 
\item How is the mechanism embedded in SUGRA? This is not an academic question
even though the physics that we are discussing is at scales far below
the Planck scale. Global supersymmetry when broken introduces a positive
vacuum energy which can only be cancelled within the context of SUGRA.
Also typically in these theories the gravitino is the lightest particle
so the whole construction is only meaningful within the locally supersymmetric
context.
\item How does one ensure that higher dimension operators, such as $\int d^{4}\theta c_{ij}\phi\bar{\phi}Q^{i}\bar{Q}^{j}/M_{P}^{2},$
where $Q^{i}$ is an MSSM field in the $i$th generation, which can
give direct coupling between the SUSY breaking sector and the visible
sector, are suppressed?
\item How is a low scale of SUSY breaking generated. Within global supersymmetry
this question has been addressed in the literature on dynamical SUSY
breaking. However as pointed out earlier this is not an issue that
can be completely settled without putting the theory within the SUGRA/string
theory framework. 
\item In string theory the couplings and mass parameters of any dynamical
symmetry breaking scheme are (functions of) string moduli which need
to be stabilized. Is the gauge mediated mechanism compatible with
this?
\item Finally do the characteristic phenomenological features of the mechanism
survive the tuning of the cosmological constant to (or close to) zero
in the quantum theory? As pointed out in \cite{deAlwis:2006nm} this
is a non-trivial issue and certainly affects gravity mediated versions
of supersymmetry breaking.
\end{itemize}
There is a large literature on mechanisms for supersymmetry breaking.
In order to generate a low scale dynamically, the phenomenon of the
generation of a mass gap in non-abelian gauge theories is invoked.
In other words the SUSY breaking scale corresponds to $\mu\sim\Lambda_{UV}e^{-8\pi^{2}/b_{0}g^{2}}$,
$ $where $g$ is the gauge coupling at some high scale $\Lambda_{UV}$
which may be the Planck scale, and $b_{0}$ is the $\beta$-function
coefficient of gauge group. However it is generally hard to get SUSY
breaking (in global SUSY) in this way. Recently it was realized that
one could nevertheless get metastable minima with broken supersymmetry
in this way \cite{Intriligator:2006dd} (for references to related
work see \cite{Intriligator:2007cp}). However a situation such as
this really needs to be addressed within supergravity as discussed
earlier. Also in string theory the gauge coupling is actually a modulus
which itself needs to be fixed. For instance in Heterotic string theory
it is proportional (at leading order) to the dilaton and in type IIB
theories where the gauge group comes from D7 branes, it is proportional
to a Kaehler modulus. Thus the phenomenon of dynamical SUSY breaking
is closely tied (in string theory) to the mechanism of moduli stabilization. 

In this work we will discuss type IIB string theory. The source of
supersymmetry breaking will be identified with the F-component of
an open string modulus acquiring a non-zero value. We will call this
field $\phi$ and write its vacuum value $\phi_{0}=M$ and its F-term
$F_{\phi}=F$.

The next issue stems from the fact that if some chiral field acquires
a non-zero F-term, then there is also a gravity mediated contribution
to SUSY breaking in the visible sector. This is governed by the value
$F/M_{P}$ while the gauge mediated contribution is governed by $F/M$.
Let $F_{0}$ be the effective F-term from all sources of SUSY breaking.
Requiring that the flavor violating contributions (coming from gravity/moduli
mediation) are small enough means for the gauge mediated ($m_{s}^{2}$)
and gravity mediated ($\Delta m_{s}^{2}$) contributions to the soft
masses (see for example \cite{Luty:2005sn})\begin{equation}
\frac{\Delta m_{s}^{2}}{m_{s}^{2}}\le10^{-3}(\frac{m_{s}}{500GeV})\label{eq:flavor}\end{equation}
where \begin{equation}
\Delta m_{s}^{2}\sim\frac{F_{0}^{2}}{M_{P}^{2}},\, m_{s}^{2}\sim\left(\frac{g^{2}}{16\pi^{2}}\right)^{2}\frac{F^{2}}{M^{2}},\label{eq:Deltamsms}\end{equation}
$g$ being the relevant gauge coupling. Requiring that the scalar
masses are not more than about $500GeV$ then gives (with $g^{2}/16\pi^{2}\sim10^{-2}$)
\begin{equation}
F\le F_{0}\le(10^{10}GeV)^{2},\, M\le10^{-3}M_{P}\frac{F^{2}}{F_{0}^{2}}\le10^{-3}M_{P}.\label{eq:FMsizes}\end{equation}
Typically in string theory the moduli acquire Planck scale vacuum
values. For instance the overall size modulus of the internal manifold
should be a factor of a few larger than the string scale for consistency
with the effective ten-dimensional low energy supergravity starting
point of KKLT type constructions. Thus clearly we need to use some
matter sector (coming from an open string sector in a type II construction)
to play the role of the field $\phi$ whose vacuum value $M$ is suppressed
relative to the Planck scale. One of the problems we hope to address
is the question of how to achieve this. However to ensure the dominance
of gauge mediated SUSY breaking and the absence of FCNC we need not
actually suppress supersymmetry breaking in the moduli sector relative
to that from the matter sector. It will be sufficient to have them
of the same order.

Finally there is the question of fine-tuning the CC. Obviously what
needs to be set to the $10^{-3}eV$ scale is the fully quantum corrected
CC which in the supergravity context one expect to be given as the
minimum of the potential with all corrections to the Kaehler potential
included. Although all our explicit calculations will be done with
the classical Kaehler potential none of the qualitative features that
we discuss will depend on its detailed form. In other words all we
are concerned with here is that the model has sufficient freedom to
achieve all fine tunings that are necessary to get a gauge mediated
model of supersymmetry breaking with the CC set to the observed scale.
We defer the discussion of what predictions of the model (if any)
survive this tuning (i.e. the question raised in \cite{deAlwis:2006nm})
to future work.

\section{SUGRA models}

We will set $M_{P}=1$ in the following. For future reference we collect
here the basic SUGRA formulae %
\footnote{See for instance Wess and Bagger\cite{Wess:1992cp} - the last three
(\ref{eq:dP}-\ref{eq:ddbarP}) may be derived using the methods of
Kaehler geometry given in Appendix C.%
} that we need (we ignore D-terms since they are not relevant for our
construction).\begin{eqnarray}
V & = & e^{K}(D_{i}WD_{\bar{j}}\bar{W}K^{i\bar{j}}-3|W|^{2})\label{eq:P}\\
\partial_{k}V & = & e^{K}(D_{l}D_{i}WD_{\bar{j}}\bar{W}K^{i\bar{j}}-2D_{k}W\bar{W})\label{eq:dP}\\
\nabla_{l}\partial_{k}V & = & e^{K}(D_{l}D_{k}D_{i}WD_{\bar{j}}\bar{W}K^{i\bar{j}}-D_{k}D_{l}W\bar{W})\label{eq:ddP}\\
\nabla_{\bar{l}}\partial_{k}V & = & e^{K}(R_{\bar{l}ki}^{\,\,\, n}D_{n}WD_{\bar{j}}\bar{WK^{i\bar{j}}}+K_{k\bar{l}}D_{i}WD_{\bar{j}}\bar{W}K^{i\bar{j}}-D_{k}WD_{\bar{l}}\bar{W}\nonumber \\
 &  & +D_{k}D_{i}WD_{\bar{l}}D_{\bar{j}}WK^{i\bar{j}}-2K_{k\bar{l}}W\bar{W}).\label{eq:ddbarP}\end{eqnarray}
In the above $\partial_{i}$ denotes differentiation with respect
to a chiral scalar $\phi^{i}$, $K_{i}=\partial_{i}K$ etc. and \begin{equation}
D_{i}W=W_{i}-K_{i}W,\,\nabla_{i}X_{j}=\partial_{i}X_{j}-\Gamma_{ij}^{k}X_{k},\,\Gamma_{ij}^{k}=K^{k\bar{l}}\partial_{i}K_{j\bar{l}},\, R_{i\bar{j}k\bar{l}}=K_{m\bar{l}}\partial_{\bar{j}}\Gamma_{ik}^{m}.\label{eq:kaehlergeom}\end{equation}
The simplest model of supersymmetry breaking is the so-called Polonyi
model \cite{Polonyi:1977pj} \begin{equation}
K=\phi\bar{\phi},\, W=\mu^{2}\phi+c.\label{eq:Polonyi}\end{equation}
The SUGRA potential associated with (\ref{eq:Polonyi}) is \begin{equation}
V=e^{\phi\bar{\phi}}[|\bar{\phi}(\mu^{2}\phi+c)+\mu^{2}|^{2}-3|\mu^{2}\phi+c|^{2}].\label{eq:Polpot}\end{equation}
This potential has no SUSY solution if $c^{2}<4\mu^{2}$ with $c,\mu^{2}$
real. If as usual $c$ is fine tuned such that the minimum is at $V_{0}=0$,
$c/\mu^{2}=2-\sqrt{3}$ and the minimum is at \begin{equation}
\phi=1-\sqrt{3},\, F\simeq\sqrt{3}\mu^{2}.\label{eq:Polmin}\end{equation}
If $\mu^{2}$ is taken to be small then the supersymmetry scale is
small. However this minimum is at large (i.e. Planck) scale and will
mean that gravity mediated SUSY breaking will dominate. To get a low
value we consider the following,\begin{equation}
K=\bar{\phi}\phi,\, W=c+\mu^{2}\phi+\frac{\nu}{2!}\phi^{2}+\frac{\lambda}{3!}\phi^{3}+\ldots,\label{eq:Polmod}\end{equation}
with\begin{equation}
c\sim\mu^{2}\sim\nu\sim\lambda\sim\ldots<<1.\label{eq:parameters}\end{equation}
As we will see in the next section this kind of model can arise naturally
in the type IIB string theory context $ $with $\phi$ being an open
string modulus. Let us look for conditions on the parameters to have
a minimum of this potential with $F_{\phi}\equiv F\ll1,\,\phi_{0}\equiv M\ll1$.
The relevant formulae are (since we have a trivial Kaehler metric
- see (\ref{eq:dP})-(\ref{eq:ddbarP})) \begin{eqnarray}
\partial_{\phi}V & = & e^{K}(D_{\phi}^{2}WD_{\bar{\phi}}\bar{W}-2D_{\phi}W\bar{W})\label{eq:Vphiderivs1}\\
\partial_{\phi}^{2}V & = & e^{K}(D_{\phi}^{3}WD_{\bar{\phi}}\bar{W}-2D_{\phi}W\bar{W})\label{eq:Vphiderivs2}\\
\partial_{\phi}\partial_{\bar{\phi}}V & = & e^{K}(D_{\phi}^{2}WD_{\bar{\phi}}^{2}\bar{W}-2W\bar{W})\label{eq:Vphiderivs3}\end{eqnarray}
For a minimum with $\phi=\phi_{0}<<1$ we have (under the conditions
(\ref{eq:parameters}) \begin{equation}
W|_{0}\sim c,\, D_{\phi}W|_{0}\sim\mu^{2},\, D_{\phi}^{2}W|_{0}\sim\nu,\, D_{\phi}^{3}W|_{0}\sim\lambda.\label{eq:Wmin}\end{equation}
The conditions for such a minimum $\partial_{\phi}V=0,$with $\partial_{\phi}\partial_{\bar{\phi}}V>0$
and $H=4\{|\partial_{\phi}\partial_{\bar{\phi}}V|^{2}-|\partial_{\phi}^{2}V|^{2}\}|_{0}>0$
then give respectively the restrictions%
\footnote{A general analysis of local stability conditions in the presence of
SUSY breaking and zero CC has been given in \cite{Gomez-Reino:2006dk}\cite{Gomez-Reino:2006wv}. %
}\begin{equation}
|\nu|\simeq2|c|,\,|\nu|^{2}>2|c|^{2},\,||\nu|^{2}-2|c|^{2}|>|\lambda\mu^{2}-|\nu c||\label{eq:parameterrelations}\end{equation}
If the first condition is a strict equality the minimum is at $\phi|_{0}=0$.
To get a minimum $\phi{}_{0}=O(\epsilon)<<1$ we thus have the fine
tuning condition \begin{equation}
|2|c|=|\nu|(1+\epsilon),\label{eq:finetuning}\end{equation}
which of course implies the two inequalities in (\ref{eq:parameterrelations}).
In addition, to avoid tachyonic messengers we need to have $\epsilon^{2}>\mu^{2}$
(so that their SUSY mass is larger than the SUSY breaking mass). In
fact from (\ref{eq:Wmin}) and (\ref{eq:FMsizes}) we see that we
need $\mu^{2}\lesssim10^{-16}$ and $\epsilon\lesssim10^{-3}$. Of
course there is also the fine-tuning condition to get a small value
of the cosmological constant:\begin{equation}
|\mu^{2}+\nu\phi_{0}+c\bar{\phi}_{0}|-\sqrt{3}|c+\mu^{2}\phi_{0}|<<O(\epsilon c)\label{eq:CCfine}\end{equation}
Thus we have two fine-tuning conditions - which can be regarded as
being for $c$ and $\nu$ (or $\mu^{2}$). Let us now add the messenger
sector $f,\tilde{f}$ (belonging for example to a $5+\bar{5}$ of
a $SU(5)$ standard model group). The modified potentials are\begin{equation}
K=\bar{\phi}\phi+\bar{f}f+\bar{\tilde{f}}\tilde{f};\, W=c+\mu^{2}\phi+\frac{\nu}{2!}\phi^{2}+\frac{\lambda}{3!}\phi^{3}+\ldots+\sigma\phi f\tilde{f},\label{eq:KWmessenger}\end{equation}
where the conditions (\ref{eq:parameters},\ref{eq:finetuning}) still
apply but $\sigma\sim O(1)$. At $f=\tilde{f}=0,\phi=\phi_{0}<<1$
we have in addition to (\ref{eq:Wmin}) the following results (with
$|$ denoting evaluation at this point).\begin{equation}
D_{f}W|=D_{f}^{2}W|=D_{\phi}D_{f}W|=D_{\phi}^{2}D_{f}W|=D_{\phi}D_{f}^{2}W|=D_{f}^{3}W|=D_{\tilde{f}}D_{f}^{2}W|=(f\leftrightarrow\tilde{f})=0\label{eq:DW=0}\end{equation}
and \begin{equation}
D_{f}D_{\tilde{f}}W|=\sigma\phi_{0},\, D_{\phi}D_{f}D_{\tilde{f}}W|=\sigma(1+\phi_{0}\bar{\phi_{0})}\label{eq:DWne0}\end{equation}
With these it can be checked that the point in question is an extremum
($\partial_{f,\tilde{f},\phi}V|=0$). Also $\partial_{\phi}\partial_{f}V|=\partial_{\phi}\partial_{\bar{f}}V|=(f\rightarrow\tilde{f})=0$
so that the $\phi$ sector mass matrices decouple from the $f,\tilde{f}$
sector. Finally in the latter sector we have,\begin{eqnarray}
\partial_{\bar{f}}\partial_{f}V| & = & \partial_{\bar{\tilde{f}}}\partial_{\tilde{f}}V|\simeq\sigma^{2}\phi_{0}^{2}+\mu^{4}-2c^{2}\label{eq:d2V}\\
\partial_{f}^{2}V| & = & \partial_{\tilde{f}}^{2}V|=\partial_{\tilde{f}}\partial_{\bar{f}}V|=0,\,\partial_{f}\partial_{\tilde{f}}V|\simeq\sigma\mu^{2}\label{eq:nondiagd2V}\end{eqnarray}
Thus if we require that (in addition to the previous conditions (\ref{eq:parameters},\ref{eq:finetuning}))
$c\sim\mu^{2}<<\phi_{0}^{2}<<1$ we see that (\ref{eq:d2V}) is positive
and the non-diagonal terms (\ref{eq:nondiagd2V}) are small compared
to the diagonal ones so that we have a true minimum.

In the next section we will consider how natural these restrictions
are in a string theory context.

\section{A string theory model}

All dynamical SUSY breaking models have a low scale generated by non-perturbative
effects. In string theory such effects are dependent on the moduli.
In the full potential for the matter ($\phi,f,\tilde{f})$ and moduli
$\Phi$ we need to find a minimum with $F_{\phi}\ne0$ and $\phi_{0}\ll1$
for some $ $matter field, while $F_{\Phi}$ need not necessarily
be much smaller, but should not be larger. As a first attempt then
we may ask whether it is possible to integrate out all the moduli
supersymmetrically, which appears to be the usual assumption in discussions
of gauge mediated supersymmetry breaking, and discuss the resulting
matter theory.

So consider the following simple toy model which is essentially the
KKLT model (with the dependence on all but one modulus - the overall
volume modulus $T$ - suppressed on the assumption that the other
moduli (and the dilaton) are stabilized at the string scale. In addition
there are matter sectors coming from the fluctuations on D3 branes
and D7 branes. The latter wrap a four-cycle in the internal manifold
and give rise to a condensing gauge group that generates the non-perturbative
contribution to the KKLT superpotential. The Kaehler potential and
superpotential are given by\begin{eqnarray}
K & = & -3\ln(T+\bar{T}-\phi\bar{\phi}-f\bar{f}-\tilde{f}\bar{\tilde{f}}),\label{eq:Kdef}\\
W & = & W_{0}+P(\phi)De^{-dT}+\sigma\phi f\tilde{f}.\label{eq:Wdef}\end{eqnarray}
This is a toy version of that which one would expect when D3 branes
are present. The chiral field $\phi$ can be regarded as the holomorphic
coordinate parametrizing the directions transverse to the four-cycle
that is wrapped by the D7 branes. The argument of the exponential
is proportional to the gauge coupling function of the gauge group
on the stack of D7 branes in the KKLT model, and the $\phi$ dependence
of the superpotential is a threshold effect. Classically in the absence
of D3 branes the gauge coupling is the $T$ modulus, but a holomorphic
dependence on the D3-brane stack position is induced in the presence
of the latter. The $\phi$ field represents the open string moduli
which determines the position of the D3-brane stack in the internal
manifold relative to the D7-brane stack. The $\phi$ dependence in
the non-perturbative term comes from the fact that the gauge coupling
is proportional to the volume and the latter is warped by the presence
of the D3-brane as observed by Baumann et al.\cite{Baumann:2006th}
(see also \cite{DeWolfe:2007hd}). $P(\phi)$ is equal to the exponential
of the holomorphic function $\zeta$ of that paper. In a particular
model for the warped throat in that work (see equation( 51)) it takes
the form\begin{equation}
P(\phi)=(1+\delta\phi)^{1/n}=(1+\frac{1}{n}\delta\phi+\frac{1-n}{2n^{2}}\delta^{2}\phi^{2}+\frac{(1-n)(1-2n)}{3!n^{3}}\delta^{3}\phi^{3}+\ldots)\label{eq:Pphi}\end{equation}
where the last equality is valid for small values of $\phi$. $\delta$
in the above parametrizes the location of the four cycle wrapped by
the D7 branes in the internal manifold and is determined (independently
of $W_{0}$ ) by the fluxes. $n$ is the number of D7 branes in the
stack.

The matter term in the argument of the logarithm in the Kaehler potential
is essentially the Kaehler potential of the CY manifold which we have
however modeled by the canonical (flat) Kaehler potential for simplicity.
We do not expect the qualitative conclusions of this work to be affected
by this simplification.

The messenger fields $f,\tilde{f}$ come from open strings and belong
to the fundamental and anti-fundamental representations of the gauge
group on the stack of D3 branes. For instance we may hope to construct
a SUSY SU(5) GUT on this stack by taking five of them in a manner
similar to the discussion (for two stacks of D7 branes) in \cite{Diaconescu:2005pc}
(see also \cite{Franco:2006ht,Garcia-Etxebarria:2006rw}). The gauge
coupling on the D3 stack is essentially fixed by the dilaton which
has been integrated out at the string scale.

If we could integrate out the modulus supersymmetrically we should
put \begin{equation}
D_{T}W=-dP(\phi)De^{-dT}-3\frac{W_{0}+P(\phi)De^{-dT}}{T+\bar{T}-\bar{\phi}\phi}=0.\label{eq:DT}\end{equation}
The number $d$ is of $O(4\pi/n)$ where $n$ \textbf{\Large }is the
number of coincident D7 branes. If the latter are of $O(1-10)$ then
$d\gtrsim O(1)$. If these conditions are satisfied and if $\phi<<1$
at the minimum then we would have $T$ determined to be the KKLT value
- i.e. \begin{equation}
dD(T+\bar{T})e^{-dT}+3(W_{0}+De^{-dT})=0\label{eq:KKLT}\end{equation}
So we need to find fluxes such that $W_{0}<<1$, in order to be consistent
with our requirement that $\Re T>1$. This determines $T=T_{0}=T_{0}(W_{0},d,D)$.
Note that this implies \begin{equation}
W_{0}\sim De^{-dT_{0}}\label{eq:W0}\end{equation}
With this we then have a model of the form discussed in the previous
section. Thus we may identify the parameters of (\ref{eq:KWmessenger})
in terms of (\ref{eq:Wdef}, \ref{eq:Pphi}) as \begin{equation}
c=W_{0}+De^{-dT_{0}},\,\mu^{2}=\frac{1}{n}\delta De^{-dT_{0}},\,\nu=\frac{1-n}{n^{2}}\delta^{2}De^{-dT_{0}},\,\lambda=\frac{(1-n)(1-2n)}{n^{3}}\delta^{3}De^{-dT_{0}}\label{eq:parameters2}\end{equation}
The suppression of the parameters in the superpotential that we need
is then obtained in a rather natural way. Let us check whether the
conditions for a minimum are satisfied. First let us ignore the fine-tuning
condition and focus on the two inequalities in (\ref{eq:parameterrelations})
imposing just the fine-tuning condition on the CC (\ref{eq:CCfine})
which for small $\phi$ gives $c\simeq\mu^{2}/\sqrt{3}$. This gives
us the conditions (for $n\ge2)$ \begin{equation}
\delta>\sqrt{\frac{2}{3}}\frac{n}{n-1},\,(n-1)(n-2)\delta^{2}+(n-1)n\frac{\delta}{\sqrt{3}}-2n^{2}>0\label{eq:deltainequality}\end{equation}
However when one imposes the the first relation in (\ref{eq:parameterrelations})
one gets the condition $\delta=2n/(\sqrt{3}(n-1)$ and this value
does not satisfy the second inequality above.

 Now one may think that the difficulty stems from the simple form
of $P(\phi)$ that we have chosen. We think this is unlikely. The
problem is that there are basically only two independent fine tunings
that can be done in this whole class of models - corresponding to
the embedding of the four cycle wrapped by the D7-branes and the value
of the flux superpotential $W_{0}$. The constraints of getting a
large value of $T_{0}$ with a small value of $\phi_{0}$ together
with a vanishingly small CC appear to be difficult to satisfy without
more freedom.

A possibly more productive strategy is perhaps to include corrections
to the Kaehler potential. But before we consider this there is another
issue that needs to be addressed. The problem is that when one requires
that $D_{T}W=0$ but $D_{\phi}W\ne0$, there is an additional fine
tuning involved. Although $D_{T}W=D_{\phi}W=0$ is a minimum (where
SUSY is of course unbroken) $D_{T}W=0,\, D_{\phi}W\ne0$ is not necessarily
even an extremum. To see this consider the extremum conditions\begin{eqnarray}
\partial_{T}V & = & e^{K}[D_{T}^{2}WD_{\bar{T}}\bar{W}K^{T\bar{T}}+D_{T}D_{\phi}WD_{\bar{\phi}}WK^{\phi\bar{\phi}}+D_{T}D_{\phi}WD_{\bar{T}}WK^{\phi\bar{T}}\nonumber \\
 &  & +D_{T}^{2}WD_{\bar{\phi}}\bar{W}K^{T\bar{\phi}}-2D_{T}W\bar{W}]\label{eq:dTV}\\
\partial_{\phi}V & = & e^{K}[D_{\phi}^{2}WD_{\bar{\phi}}\bar{W}K^{\phi\bar{\phi}}+D_{\phi}D_{T}WD_{\bar{\phi}}WK^{T\bar{\phi}}+D_{\phi}D_{T}WD_{\bar{T}}WK^{T\bar{T}}\nonumber \\
 &  & +D_{\phi}^{2}WD_{\bar{T}}\bar{W}K^{\phi\bar{T}}-2D_{\phi}W\bar{W}].\label{eq:dphiV}\end{eqnarray}
We see that to have $D_{T}W=0$ but $D_{\phi}W\ne0$ (so as to reproduce
the analysis of section three) we would need additional fine tuning
of parameters to set for instance $ $$D_{T}D_{\phi}W=0$. However
we have used up the freedom to fine tune.

Let us revisit the issue by first looking at SUSY minima which are
of course generically AdS. We continue to ignore the messenger sector
to focus on the $\phi,T$ sector. For SUSY solutions we have,\begin{eqnarray}
D_{T}W & = & -dPDe^{-dT}-\frac{3W}{T+\bar{T}-\phi\bar{\phi}}=0\label{eq:susyT}\\
D_{\phi}W & = & P'(\phi)De^{-dT}+\frac{3\bar{\phi}}{T+\bar{T}+\phi\bar{\phi}}W=0\label{eq:susyphi}\end{eqnarray}
With $P(\phi)=(1+\delta\phi)^{1/n}$ as before. Note that we can eliminate
$T$ from these two equations and get \begin{equation}
\bar{\phi}_{0}=-\frac{1}{nd(1+\delta\phi_{0})}\sim O(1/4\pi),\,{\rm for}\, nd\sim4\pi\label{eq:phisusy}\end{equation}
The SUSY mass matrices are (from (\ref{eq:ddbarP},\ref{eq:ddP})
with $D_{T,\phi}W=0$) are \begin{eqnarray}
\partial_{\bar{T}}\partial_{T}V|_{0} & = & e^{K_{0}}[D_{T}^{2}WD_{\bar{T}}^{2}\bar{W}K^{T\bar{T}}+D_{T}D_{\phi}WD_{\bar{T}}^{2}{}_{\bar{\phi}}\bar{W}K^{\phi\bar{\phi}}+2\Re D_{T}^{2}WD_{\bar{T}}^{2}{}_{\bar{\phi}}\bar{W}K^{T\bar{\phi}}\label{eq:TTbar}\\
 &  & -2K_{T\bar{T}}|W|^{2}]\nonumber \\
\partial_{\bar{\phi}}\partial_{\phi}V|_{0} & = & e^{K_{0}}[D_{\phi}^{2}WD_{\bar{\phi}}^{2}\bar{W}K^{\phi\bar{\phi}}+D_{T}^{2}{}_{\phi}WD_{\bar{T}}^{2}{}_{\bar{\phi}}\bar{W}K^{T\bar{T}}+2\Re D_{\phi}^{2}WD_{\bar{T}}^{2}{}_{\bar{\phi}}\bar{W}K^{\phi\bar{T}}\label{eq:phiphibar}\\
 &  & -2K_{\phi\bar{\phi}}|W|^{2}]\nonumber \\
\partial_{\bar{T}}\partial_{\phi}V|_{0} & = & e^{K_{0}}[D_{\phi T}^{2}WD_{\bar{T}}^{2}\bar{W}K^{T\bar{T}}+D_{\phi}^{2}WD_{\bar{T}}^{2}{}_{\bar{\phi}}\bar{W}K^{\phi\bar{\phi}}+2\Re D_{\phi}^{2}{}_{T}WD_{\bar{T}}^{2}{}_{\bar{\phi}}\bar{W}K^{T\bar{\phi}}\label{eq:Tbarphi}\\
 &  & -2K_{\phi\bar{T}}|W|^{2}]\nonumber \\
\partial_{T}^{2}V & |_{0}= & -e^{K_{0}}D_{T}^{2}W\bar{W},\,\partial_{\phi}^{2}V=-e^{K_{0}}D_{\phi}^{2}W\bar{W},\,\partial_{T}^{2}{}_{\phi}V=-e^{K_{0}}D_{T}^{2}{}_{\phi}W\bar{W}\label{eq:TTetc}\end{eqnarray}
Here the right hand sides are to be evaluated at the extremum, i.e.
the solution to (\ref{eq:susyT},\ref{eq:susyphi}). Now as is well-known
although at a Minkowski minimum (i.e. one with $W|_{0}$ fine tuned
to be zero) these mass matrices are positive definite, this is not
necessarily the case for an AdS minimum. For our purposes, where we
want to integrate out one of the fields (i.e. $T$) to get a theory
for the other with hopefully a non-AdS (meta-stable) minimum, we do
need a positive definite mass matrix as a starting point. Thus we
need to ensure that the first two in the above set are positive while
the others are subdominant compared to these two. Evaluating the various
derivatives we have \begin{equation}
D_{T}^{2}W\sim O(d^{2}e^{-dT_{0}}),\, D_{\phi}^{2}W\sim O(\frac{\delta^{2}}{n}e^{-dT_{0}})\sim O(de^{-dT}),\, D_{\phi}D_{T}W\sim O(\frac{d\delta}{n}e^{-dT_{0}}),\, W\sim O(e^{-dT_{0}})\label{eq:Od^2W}\end{equation}
Then we see that by choosing $\delta>d$ (recall that $d\sim4\pi/n$)
we have $|D_{T}^{2}W|,\,|D_{\phi}^{2}W|>|D_{\phi}D_{T}W|$. This would
guarantee that the $T,\phi$ mass matrix has positive eigenvalues
(since the traces in each subsector is positive and the off-diagonal
terms are smaller than the diagonal ones). However the problem is
that the mass for the mostly $T$ field is of the same order as that
of the (mostly) $\phi$ field. They are both suppressed by the exponential
factor. This is of course in contrast to the masses of the other moduli
and the dilaton which get masses from the fluxes rather than from
the non-perturbative term. Thus integrating out $T$ at this supersymmetric
point and moving in the $\phi$ direction to find a SUSY breaking
point while assuming that the dynamics of $T$ is frozen at the point
$D_{T}W=0$ is not justified.

What one needs to do then is to find a point $\partial_{T}V=\partial_{\phi}V=0$
with $T_{0}\gtrsim1,\, D_{\phi}W\ne0,\,\phi_{0}<<1,\, V_{0}=0$. The
last two conditions are then fine tuning conditions. As for the F
term of the T-modulus we do not need to require that it be zero or
even much smaller than $F_{\phi}$. All that is necessary is that
it should not be larger as we discussed earlier. This is a mild condition
on the flux parameters $c,$$\delta$ and should not be hard to realize.
However as we argued earllier the fine-tuning that is required to
get a small value of $\phi_{0}$ is not compatible with the stability
conditions in this class of models. 

Nevertheless there appears to be a possibility of obtaining the desired
result with just the fine tuning of the CC by a small modification
of this model. This involves including the effects of integrating
out string/KK states. In fact this way one gets a model like the one
studied by Kitano\cite{Kitano:2006wz} (see also \cite{Abe:2006xp}
and \cite{Kallosh:2006dv}). Let us first review the relevant part
of \cite{Kitano:2006wz}. The Kaehler and superpotentials are taken
to be \begin{equation}
K=\phi\bar{\phi}+\frac{(\phi\bar{\phi})^{2}}{\Lambda^{2}}+\ldots,\, W=c+\mu^{2}\phi+\ldots\label{eq:kitano}\end{equation}
The higher order terms in $K$ may arise from integrating out states
at the scale $\Lambda<1$ (in Planck units). Also $c,\mu^{2}<<\Lambda$.
Then the minimum (once we adjust $c$ so that the potential is zero
at the minimum) is at\[
|\phi_{0}|\simeq\frac{\Lambda^{2}}{2\sqrt{3}}\ll1\]
$ $ Thus we can get a minimum with a low value of $\phi$ without
fine-tuning if there is another (high) scale that is still well below
the Planck scale. In string theory this may come from integrating
out Kaluza-Klein states or string states. Here we identify $\Lambda$
with the string scale and regard the extra terms as coming from integrating
out string states. However the string scale is in fact dependent (in
our Einstein frame considerations) on the size of the internal manifold.
(It also depends on the dilaton which we assume has been stabilized
such that the string coupling $g_{s}\sim O(1)$ and on the warping
due to the fluxes and branes and which we also take to be of $O(1)$
for simplicity). Then in our type IIB context (with $M_{P}=1$) \begin{equation}
M_{s}^{2}\sim T_{R}^{-3/2}<1,\label{eq:Lambda}\end{equation}
where of course the last inequality is valid once the volume modulus
has been stabilized at a large value. Then for small values of $\phi\bar{\phi}$,
expanding around the value of $T_{R}$ at the minimum and rescaling
$\phi$\begin{equation}
K=-3\ln(T+\bar{T})+\phi\bar{\phi}+\frac{1}{6}(\phi\bar{\phi})^{2}+2T_{0R}^{4}(\phi\bar{\phi})^{2}+\ldots\label{eq:Kexpansion}\end{equation}
Now we can repeat the previous analysis. We use the freedom of tuning
$c$ or (equivalently $W_{0}$) to set the cosmological constant to
zero at the minimum of the potential $\partial_{T}V=\partial_{\phi}V=0$
where the values at the minimum are themselves functions of the parameters
$W_{0},\delta,n$. Note that with a sufficiently complicated internal
manifold (say a Calabi-Yau with $O(10^{2})$ complex structure moduli)
one can imagine making the first two of these parameters almost continuous
(thus satisfying these two equations) but $n$ can only take a limited
number of integral values. This latter freedom can still be used to
make sure that the minimum is at a large value of $T_{R}$. However
now we do not need to fine tune in order to get a small value for
$\phi$. By choosing $\delta>d$ (see (\ref{eq:Od^2W})) we see that
in (\ref{eq:dphiV}) we can ignore the second and third terms in the
square bracket compared to the other terms. The fourth term would
be small for small $\phi$ since $K_{\phi\bar{T}}\sim O(\phi)$. $\phi$
is then effectively determined by the first and the last terms (i.e.
as if $T$ is fixed). Thus it is determined by the largest higher
dimension operator i.e. the last term in (\ref{eq:Kexpansion}) as
was the case in \cite{Kitano:2006wz} so that \begin{equation}
|\phi|\sim\Lambda^{2}\sim T_{0R}^{-4}\ll1.\label{eq:phimin}\end{equation}

Also we do not need to ensure that the SUSY breaking is entirely from
the $\phi$ field. All that one needs to ensure is that the F-term
from the modulus is not larger than that from the $\phi$ field and
this is a relatively mild condition which should not be hard to satisfy.
Strictly speaking we should do an  analysis of the multidimensional
potential (i.e with 6 real dimensions corresponding to $\phi,f,T$
) but our analysis at least makes it plausible that it is possible
in this modified KKLT model to obtain a SUSY breaking minimum with
$F_{\phi}\sim F_{T},\,\phi_{0}\ll1$ so that gauge mediated mechanism
can be the dominant mode of supersymmetry breaking in the MSSM sector.

Let us make a few numerical checks. Assuming that the fluxes and the
gauge theory on the D7 branes can be chosen such that $T_{R}\gtrsim20$,
$M_{s}\sim T^{-3/4}\lesssim10^{-1}$, corresponding to a string scale
$\lesssim10^{17}GeV$. Then $|F_{0}|\sim e^{-4\pi T_{R}/n}\le6\times10^{-19}$
for $n\le6$ and $M\equiv|\phi_{0}|\sim\Lambda^{2}\sim T_{R}^{-4}<10^{-4}$
so that the criteria in (\ref{eq:FMsizes}) are satisfied. The vacuum
in which gauge mediated SUSY breaking is realized is as usual metastable
since generically there will be both supersymmetric (see (\ref{eq:susyphi}))
and non-supersymmetric vacua with negative energy. However in SUGRA
(as argued in \cite{Weinberg:1982id}) one expects the decay rates
to these vacua to be strongly suppressed.

\section{Conclusions}

The question we have addressed here is whether gauge mediated supersymmetry
breaking is achievable within string theory with all moduli stabilized.
Any theory that has supersymmetry breaking must of course confront
the fine-tuning of the cosmological constant and this is achieved
in a SUGRA model by fine-tuning a constant in the superpotential.
In string theory the constant is related to the flux background in
the extra dimensions and the fine tuning is discrete, but with a sufficiently
complicated internal manifold this can be achieved. However to get
a low scale of SUSY breaking so that SUSY mass splittings are at the
TeV scale, one needs an exponential suppression compared to the Planck
scale. This is usually achieved in global considerations of dynamical
SUSY breaking by having a gauge theory that develops a mass gap. Here
the same mechanism \cite{Kachru:2003aw} that helps to stabilize the
volume modulus $T$ (or more generally the Kaehler moduli) is used
as the mechanism for getting a suppressed supersymmetry breaking scale.
In addition, to have gauge mediated SUSY breaking the matter field
breaking supersymmetry must have a small (compared to the Planck scale)
expectation value. In order to do this however we need to include
terms that would arise from integrating out string (or KK) modes.
We argued that the SUSY breaking field can be stabilized at the string
scale, which for large volume compactification can be much less than
the Planck scale, and that then it is possible to get a model with
gauge mediated SUSY breaking that would only require the single fine-tuning
of the CC. 

Nevertheless it should be emphasized that the main message of this
paper is not so much the fact that it is possible to get a gauge mediated
SUSY breaking model in the context of a string theory model. It is
rather that discussions of the former cannot be decoupled from the
latter in theories (like the KKLT one) where some of the moduli require
non-perturbative effects for stabilization, so that it is not really
posssible to get a large mass hierarchy between these moduli and the
SUSY breaking sector. So a viable theory of gauge mediated supersymmetry
breaking (at least in the framework of type IIB string theory) needs
to address together, both the stabilization of  the Kaehler moduli,
and the open string modulus that is responsible for the SUSY breaking.

\section{Acknowledgments}

I wish to thank Ramy Brustein for very useful comments on the manuscript
and Oliver DeWolfe and Ben Shlaer for discussions. This research is
supported in part by the United States Department of Energy under
grant DE-FG02-91-ER-40672.

\bibliographystyle{apsrev}
\bibliography{myrefs}

\end{document}